\documentclass[prl,superscriptaddress,showpacs,longbibliography,reprint]{revtex4-2} 

\usepackage{hyperref}
\usepackage{color}

\usepackage[usenames,dvipsnames]{xcolor}
\usepackage{amsmath}
\usepackage{amssymb}
\usepackage{graphicx}
\graphicspath{{graphics/}}
\usepackage{epsfig}
\usepackage{dcolumn}
\usepackage{bm}
\usepackage{mathrsfs}
\usepackage{multirow}
\usepackage[all]{xy}
\usepackage{pbox}
\usepackage{lipsum}
\usepackage{verbatim}
\usepackage{braket}
\usepackage{dsfont}
\usepackage{array}
\usepackage{makecell}
\usepackage{tabularx}
\usepackage{isotope}

\def\(({\left(}
\def\)){\right)}
\def\[[{\left[}
\def\]]{\right]}

\usepackage{graphicx}
\graphicspath{{./Figures/}}

\usepackage{color,soul}

\begin{document}
\title{Nonequilibrium dark space phase transition}

\author{Federico Carollo}
\email[Corresponding Author: ]{federico.carollo@uni-tuebingen.de}
\affiliation{Institut f\"ur Theoretische Physik, Eberhard Karls Universit\"at T\"ubingen, Auf der Morgenstelle 14, 72076 Tübingen, Germany}
\author{Igor Lesanovsky}
\affiliation{Institut f\"ur Theoretische Physik, Eberhard Karls Universit\"at T\"ubingen, Auf der Morgenstelle 14, 72076 Tübingen, Germany}
\affiliation{School of Physics and Astronomy and Centre for the Mathematics and Theoretical Physics of Quantum Non-Equilibrium Systems, The University of Nottingham, Nottingham, NG7 2RD, United Kingdom}
\begin{abstract}
We introduce the concept of dark space phase transition, which may occur in open many-body quantum systems where irreversible decay, interactions and quantum interference compete. Our study is based on a quantum many-body model, that is inspired by classical nonequilibrium processes which feature phase transitions into an absorbing state, such as epidemic spreading. The possibility for different dynamical paths to interfere quantum mechanically results in collective dynamical behavior without classical counterpart. We identify two competing dark states, a trivial one corresponding to a classical absorbing state and an emergent one which is quantum coherent. We establish a nonequilibrium phase transition within this dark space that features a phenomenology which cannot be encountered in classical systems. Such emergent two-dimensional dark space may find technological applications, e.g. for the collective encoding of a quantum information.
\end{abstract}

\maketitle
A dark (or absorbing) state is a non-fluctuating state that once it is reached during the course of a time-evolution it cannot be left. Dynamical systems that possess a dark state can display complex nonequilibrium behavior and universal dynamical scaling, even in low dimensions \cite{lubeck2005,henkel2008,hinrichsen2000}. Remarkably, many real-world processes actually feature such dark state, as, for instance, the epidemic spreading of a virus among a population \cite{mollison1977,grassberger1983}: for sufficiently low infection rate the population reaches a dark state, where all units are healthy and the virus is eradicated. However, when the infection rate is increased a stationary state phase transition to a fluctuating phase can take place. Here the virus becomes endemic and an extensive number of units remains infected. Interestingly, also dissipative quantum processes can feature dark states and allow to explore related concepts and phenomena in an entirely different setting. However, the phenomenology of systems studied so far \cite{marcuzzi2016,gutierrez2017,roscher2018,carollo2019,gillman2019,helmrich2020,jo2021,nigmatullin2021} is closely related to that of classical processes.
\begin{figure}[t]
    \centering
    \includegraphics[width=\linewidth]{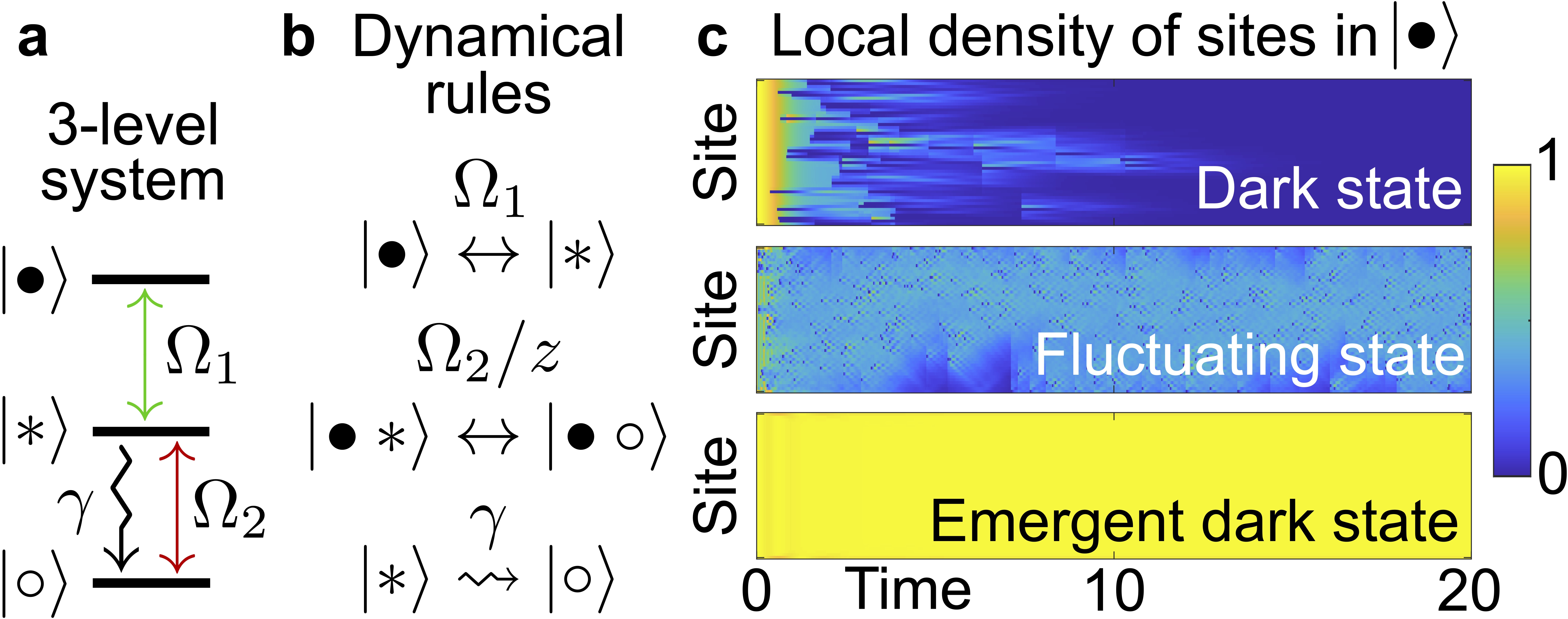}
    \caption{\textbf{Dark state phase transition}. (a) Three-level quantum system with basis states $\ket{\bullet}$,$\ket{*}$ representing contagious and non contagious infected units, and the healthy state $\ket{\circ}$. (b) The dynamics consists of coherent transitions between infected states $\ket{\bullet}\leftrightarrow\ket{*}$ (rate $\Omega_{1}$) and between states $\ket{\circ}\leftrightarrow\ket{*}$. This (infection) process must be facilitated by the presence of a contagious neighbor. For each contagious neighbor, the rate of the process is enhanced by a factor $\Omega_{2}/z$, where $z$ is the coordination number of the lattice. (c) Illustrative trajectories for a $1D$ quantum system with $50$ sites. Top: Approach to the dark state $\ket{\rm D}$ for the model depicted in (a). Middle: Fluctuating phase which typically emerges in classical and quantum models with absorbing states. The example shown is for the quantum contact process \cite{carollo2019}. Bottom: Emergence of the dark state $\ket{\rm D_e}$ for the model in depicted in (a).}
    \label{fig:fig1}
\end{figure}

In this paper, we report analytical and numerical evidence for the existence of a novel type of dark state phase transition, which has no classical counterpart as it crucially relies on quantum interference. To illustrate this new phenomenology we utilize a quantum many-body system, composed of $N$ units which can be found in three different states [shown in Fig. \ref{fig:fig1}(a)]. Using the analogy of epidemic spreading, one state represents a healthy unit, denoted as $\ket{\circ}$. The second state, $\ket{*}$, represents instead an infected but not contagious unit, while the third, $\ket{\bullet}$, represents an infected unit which is also contagious and can thus spread the virus. The dynamics of these units is subject to a classical process --- the recovery process --- consisting of transitions from state $\ket{*}$ to the $\ket{\circ}$, which competes with two other processes that are quantum coherent. The first one connects the contagious and not contagious states. The second one can be regarded as a quantum analogue of an infection process: coherent transitions between the state $\ket{\circ}$ and $\ket{*}$ take place, provided that at least one of the neighbors of the unit is in the contagious state [see Fig. \ref{fig:fig1}(b)]. 

According to these dynamical rules, the state with all healthy units, $\ket{\rm D}$, is an exact dark state for any system size $N$. Indeed, such state has no contagious unit that may activate the spreading of the infection [see Fig. \ref{fig:fig1}(c)]. Typically, for both classical and quantum  dark state phase transitions \cite{lubeck2005,henkel2008,hinrichsen2000,marcuzzi2016,gutierrez2017,roscher2018,carollo2019,gillman2019,jo2021}, one observes, for increasing infection rate, the emergence of a second steady state with finite density $\varrho_\bullet$ of contagious sites. This state exhibits dynamical fluctuations [c.f. Fig. \ref{fig:fig1}(c)]. However, the model depicted in Fig. \ref{fig:fig1}(a) displays behavior which is markedly different: the second stationary state is an emergent dark state $\ket{\rm D_e}$, which shows no fluctuations [c.f. Fig. \ref{fig:fig1}(c)] and --- contrary to the state $\ket{\rm D}$ --- it is a genuine quantum state characterized by units being in a superposition of contagious and healthy states. Beyond the novel phase transition phenomenology, the emergent dark manifold could potentially play a role in quantum information. The two dark states encode a qubit state $\ket{\Psi}=\alpha\ket{\rm D}+\beta\ket{\rm D_e}$ and the robustness of these two non-fluctuating stationary phases could provide an efficient self-correcting mechanism. 
\vspace{10pt}

\noindent {\bf The model.---} The elementary processes of the considered many-body quantum system are shown in Fig. \ref{fig:fig1}(a). It consists of $N$ three-level units, which are arranged in a $D$-dimensional lattice and follow a stochastic Markovian evolution \cite{gardiner2004} whose dynamical realizations (quantum trajectories) are governed by a random process through a stochastic Schr\"odinger equation \cite{plenio1998,gardiner2004}. It is convenient to first introduce the deterministic time-evolution of the quantum state averaged over all trajectories. This state is described, at any time $t$, by the density operator $\rho_t$ which obeys the quantum master equation \cite{lindblad1976,gorini1976}
\begin{eqnarray}
\frac{d}{dt}{\rho}_t=\mathcal{L}[\rho_t]=-i[H,\rho_t]+\mathcal{D}\left[\rho_t\right].
\label{Lindblad}
\end{eqnarray}
The super-operator
\begin{equation}
\mathcal{D}[\rho]=\gamma\sum_{k=1}^L\left(J^{(k)}_-\rho J^{ (k)}_+-\frac{1}{2}\left\{J^{ (k)}_+ J^{(k)}_-,\rho \right\}\right)\, 
\label{dissipator}
\end{equation}
is the so-called dissipator. In our case, it accounts for the classical (irreversible) transitions from the infected state $\ket{*}$ to the healthy one $\ket{\circ}$ [c.f. Fig. \ref{fig:fig1}(b)], and the jump operator $J_-=\ket{\circ}\!\bra{*}$ (with $J_+=J_-^\dagger$) implements the desired transition. The superscript $k$ indicates the site onto which the operator acts, while $\gamma$ is the rate at which the transition occurs. 

The coherent dynamics, see also Fig. \ref{fig:fig1}(b), is governed by the Hamiltonian
\begin{equation}
H=\sum_k \left[\Omega_{1} \lambda^{(k)}_1+\Omega_{2}\Pi^k_{\bullet}\lambda^{(k)}_6\right]\, ,
\label{Hamiltonian}
\end{equation}
where we have defined the (Gell-Mann matrices) $\lambda_1=\ket{\bullet}\!\bra{*}+{\rm h.c.}$ and $\lambda_{6}=\ket{*}\!\bra{\circ}+{\rm h.c.}$ The operator  $\Pi_{\bullet}^k$ implements the dynamical constraint required for the infection process [c.f. Fig.~\ref{fig:fig1}(b)] and its precise structure depends on the lattice geometry. 

The quantum master equation \eqref{Lindblad} governs the dynamics of the average (in general mixed) state. At the level of quantum trajectories \cite{plenio1998}, the quantum state is pure for all times but follows a piece-wise deterministic evolution: the state evolves according to the deterministic (non-linear) equation 
\begin{equation}
\frac{d}{dt}\ket{\psi_t}=\left[-{\rm i}H_{\rm eff}+{\rm i}\bra{\psi_t}H_{\rm eff}\ket{\psi_t}\right]\ket{\psi_t}\, ,
\label{H_eff_evo}
\end{equation}
where $H_{\rm eff}=H-\frac{{\rm i}  \gamma  }{2}\sum_k J^{(k)}_+ J^{(k)}_-$ is the (non-Hermitean) effective Hamiltonian. However, at random times, a transition $\ket{*}\to \ket{\circ}$ occurs at a random site $k$, resulting in an abrupt jump of the quantum state $\ket{\psi_t}\to J_-^{(k)}\ket{\psi_t}$. This means that the $k$-th unit has healed (it can get infected again). More precisely, the transition rate for site $k$ is given by $w_t^k=\gamma \langle n_{*}^{(k)}\rangle_t$, with $n_{*}=\ket{*}\!\bra{*}$ and $\langle \cdot\rangle_t$ denoting the quantum expectation value with respect to the state $\ket{\psi_t}$. After a jump, the dynamics under Eq. \eqref{H_eff_evo} resumes until the next jump occurs. 

The effective Hamiltonian has complex eigenvalues $c_i$, whose imaginary part $r_i=-{\rm Im}(c_i)$ is (half of) the escape rate from the associated eigenstate. 
\begin{figure*}[t]
    \centering
    \includegraphics[width=\linewidth]{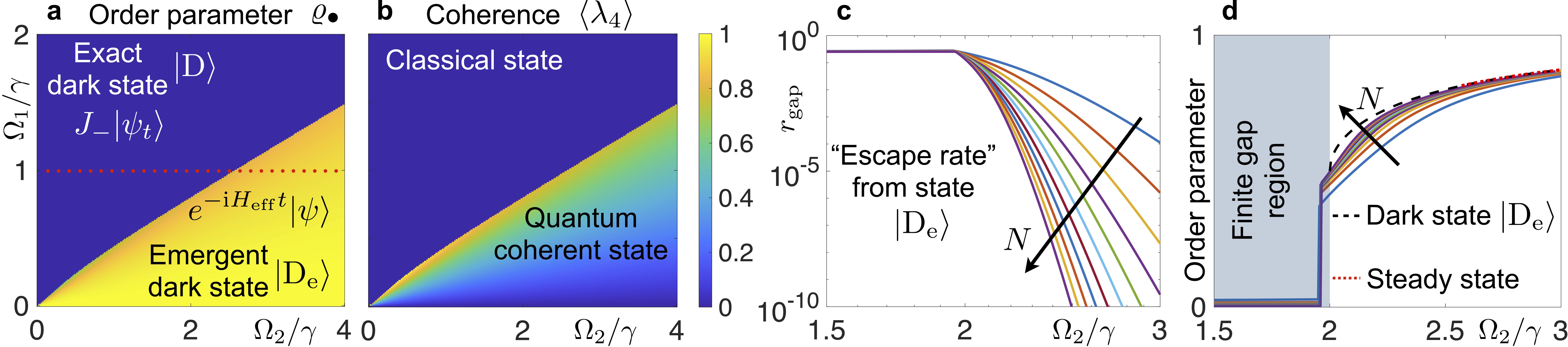}
    \caption{\textbf{Infinite-dimensional lattice}. (a) Stationary behavior of the density of contagious sites $\varrho_\bullet$, which is an order parameter for the dark space phase transition. One phase is dominated by the classical jumps $\ket{*}\to \ket{\circ}$, bringing the system towards the state $\ket{\rm D}$. The other is dominated by the ``no jump" dynamics under $H_{\rm eff}$, which drives the system towards the dark state $\ket{\rm D_e}$. (b) Finite expectation values of $\lambda_4$ demonstrate that $\ket{\rm D_e}$ features coherence between states $\ket{\bullet}$ and $\ket{\circ}$. (c) Log-log plot of the gap ($r_{\rm gap}$) of $H_{\rm eff}$ for $\Omega_1=\gamma$ as a function of $\Omega_2/\gamma$.  This quantity is half the escape rate from the dark state $\ket{\rm D_e}$ and rapidly vanishes for increasing $N$, when $\Omega_2>2\Omega_1$. (d) Average density $\varrho_\bullet$ for the eigenvector of $H_{\rm eff}$ associated with the gap. This shows a phase transition, from a region where the density is zero to a region where the density is finite. For increasing $N$, the numerical results approach the analytic prediction. Highlighted in red, the region where the average dynamics starting from $\ket{\rm U}$ approaches the dark state $\ket{\rm D_e}$. The finite $N$ curves are for $N=20,30,40,\dots 120$.  }
    \label{fig:fig2}
\end{figure*}
The survival probability of a general state $\ket{\psi}$, i.e. its probability to evolve according to Eq. \eqref{H_eff_evo} for a time $t$ without jumps, is $s_t(\ket{\psi})=\left\|   e^{-{\rm i}  H_{\rm eff} t  } \ket{\psi} \right\|^2$. The state with all healthy units, $\ket{\rm D}=\bigotimes_{k=1}^N\ket{\circ}^{(k)}$, cannot be left once reached dynamically. Indeed, we have $s_t(\ket{\rm D})=1$, meaning that the state will experience zero jumps with probability $1$. Mathematically, this is a consequence of $\ket{\rm D}$ being an eigenstate of $H_{\rm eff}$, associated with the eigenvalue $0$. Thus, $\ket{\rm D}$ has escape rate $r_{\rm min}=0$ and is  invariant under the evolution in Eq. \eqref{H_eff_evo}. It is an exact dark state for any $N$. In what follows, we show that, for $N\to\infty$ and for sufficiently large $\Omega_2$, $H_{\rm eff}$ develops a second smallest escape rate $r_{\rm gap}$ (the ``gap" of $H_{\rm eff}$), such that $r_{\rm gap}\to0$. This vanishing escape rate is related to an emergent (second) dark state $\ket{\rm D_e}$, which, has a finite density $\varrho_{\bullet}$ [see Fig. \ref{fig:fig1}(c)]. This determines a phase transition between dark states in the steady state of the average quantum dynamics.  
\vspace{10pt}

\noindent {\bf Infinite dimension.---} To establish the existence of the nonequilibrium dark space phase transition, we first focus on the limit of an infinite-dimensional lattice. Here, each site has all the others as neighbors and the constraint can be written as 
\begin{equation}
\Pi_\bullet^k=\frac{1}{N-1}\sum_{h, h\neq k}^N n_\bullet^{(h)}\approx \frac{1}{N}\sum_{h=1}^N n_\bullet^{(h)}\, , 
\label{constrain-infinite}
\end{equation}
with $n_\bullet=\ket{\bullet}\!\bra{\bullet}$. The constraint thus requires a finite density of contagious sites. The resulting open quantum dynamics in Eq. \eqref{Lindblad} can be exactly solved in the thermodynamic limit $N\to\infty$  \cite{benatti2018,carollo2020}. We can thus investigate both dynamical and stationary values for the density of sites, as well as the coherence measured by the operator $\lambda_4=\ket{\bullet}\!\bra{\circ}+{\rm h.c.}$ (see Methods). Throughout, we consider the state in which all units are contagious, $\ket{\rm U}=\bigotimes_{k=1}^N\ket{\bullet}^{(k)}$, as initial state. For weak infection rate $\Omega_{2}$, the dynamics features a unique steady state --- the dark state $\ket{\rm D}$. As $\Omega_2$ increases, two further stationary states emerge which contain a finite density of contagious sites. To understand whether these are dynamically relevant we have numerically integrated the equations of motion and found that only one of them can be approached dynamically [as shown in Fig. \ref{fig:fig2}(a)]. This is indeed the emergent pure dark state $\ket{\rm D_e}$. 

It is interesting to investigate how the two dark states, $\ket{\rm D}$ and $\ket{\rm D_e}$, which constitute nonequilibrium phases are  approached from the initial state. Approaching state $\ket{\rm D}$, the dynamics is dominated by quantum jumps, which take the system towards this classical dark state. The approach to state $\ket{\rm D_e}$ is instead dominated by the no-jump evolution under the effective Hamiltonian $H_{\rm eff}$. In this regime, even if quantum jumps occur, the deterministic dynamics in Eq. \eqref{H_eff_evo} prevails and eventually brings the system towards the emergent dark state $\ket{\rm D_e}$, which features quantum superposition of contagious and healthy states, see Fig. \ref{fig:fig2}(b). The emergence of the second dark state implies that $H_{\rm eff}\ket{\rm D_e}=c_2 \ket{\rm D_e}$, with $r_2=r_{\rm gap}=-{\rm Im}\, c_2\to0$ in the thermodynamic limit. This indeed means that the dynamics in Eq. \eqref{H_eff_evo} has also the state $\ket{\rm D_e}$ as a fixed point, and that $s_t(\ket{\rm D_e})=1$, so that this state is protected against quantum jumps. 

To verify this picture, we have diagonalized $H_{\rm eff}$ and studied its spectrum in the sector of fully symmetric many-body states (see Methods and Ref. \cite{SM}). In Fig. \ref{fig:fig2}(c), we see that there is a range of $\Omega_2$-values where the gap, $r_{\rm gap}$, remains finite. Here the system has $\ket{\rm D}$ as the sole steady state. For larger $\Omega_2$, the gap decreases with the system size with a trend that indicates a rapid convergence to zero for $N\to\infty$. We note that at the critical point the gap decays with a power-law $r_{\rm gap}\propto N^{-z}$, with $z\approx 0.3$ \cite{SM}. As shown in Fig. \ref{fig:fig2}(d), the first ``excited" state of $H_{\rm eff}$ develops, in this region, a finite density of contagious sites $\varrho_\bullet$ which tends to the steady state prediction obtained for the Lindblad dynamics. This eigenvector is the emergent dark state $\ket{\rm D_e}$.
\vspace{10pt}

\begin{figure}[t]
    \centering
    \includegraphics[width=\linewidth]{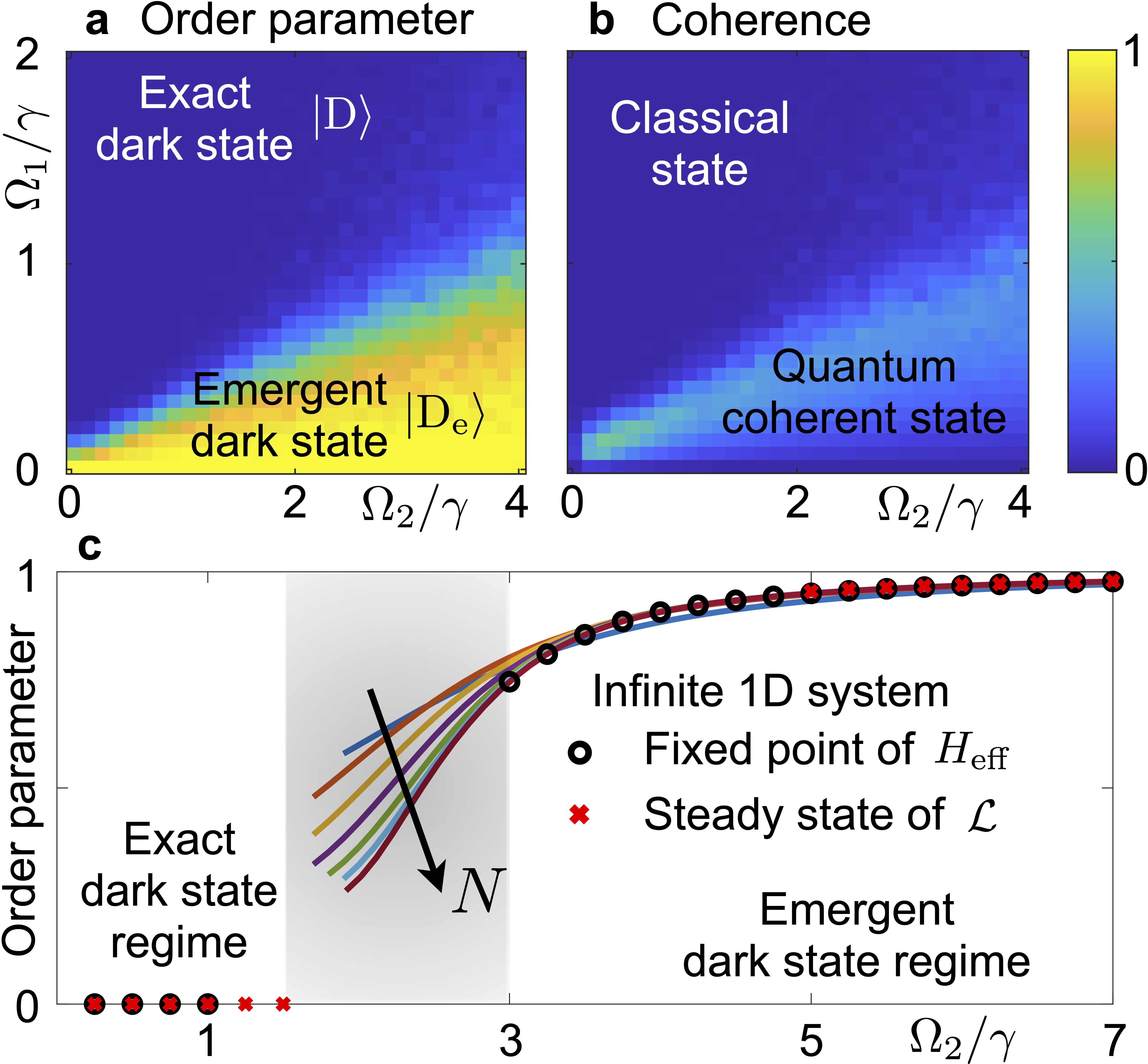}
    \caption{\textbf{One-dimensional lattice}. (a-b) Simulations of quantum jump trajectories, for a $1D$ system with $N=6$. Each data point is obtained by averaging over $100$ trajectories. The panels show the behavior of the density of contagious sites $\varrho_\bullet$ and of the coherence $\braket{\lambda_4}$, for $t=20/\gamma$. Two different phases emerge: one is the dark state $\ket{\rm D}$, while, in the other the system approaches the emergent dark state $\ket{\rm D_e}$. (c) Order parameter $\varrho_\bullet$ as a function of $\Omega_2$, for $\Omega_1=\gamma$ in the emergent dark state. For finite systems, this is estimated by looking at properties of the state associated with the gap of the effective Hamiltonian (solid lines for $N=3,4,\dots 9$). For infinite systems, we exploit matrix product state methods to target the fixed point of the dynamics in Eq.\eqref{H_eff_evo} (black circles). This shows a transition from $\ket{\rm D}$ to $\ket{\rm D_e}$. We have further obtained values of the density $\varrho_\bullet$ in the steady state of the Lindblad dynamics Eq.\eqref{Lindblad} (red crosses), which are in agreement with the prediction made for $H_{\rm eff}$. In the shaded region, our MPS algorithms did not converge. }
    \label{fig:fig3}
\end{figure}

\noindent {\bf One-dimensional lattice.---} We now focus on a $1D$ lattice, where a given site $k$ has only two neighbors and the constraint $\Pi_\bullet^k$ reads as
\begin{equation}
\Pi_\bullet^k=\frac{n_\bullet^{(k-1)}+n_\bullet^{(k+1)}}{2}\, .
\label{constraint-1D}
\end{equation}
For this setting, no analytical solution is possible. We therefore rely on extensive numerical investigations to show the emergence of the dark state $\ket{\rm D_e}$ (see Methods).

In Fig. \ref{fig:fig3}(a-b) we show results obtained from sampling quantum trajectories for a few-body system up to a finite time $t$. The ``phase diagram" displays a behavior similar to what we observed for the infinite-dimensional lattice. However, in this case, the appearance of the dark state $\ket{\rm D_e}$ is only possible for a transient period of time, given that, for finite $N$, the unique steady state is the exact dark state $\ket{\rm D}$. Representative trajectories for larger system sizes [see top and bottom trajectories in Fig. \ref{fig:fig1}(c)] show clearly how one phase is reached when quantum jumps are dominating the dynamics, while the other one is approached when the system state is driven towards the emergent dark state by the effective Hamiltonian. 

To talk about a proper nonequilibrium phase transition in the quantum system, we need to address the thermodynamic limit of an infinitely long chain. We do this by exploiting methods based on matrix product states (MPSs) \cite{vidal2003,Vidal2004,vidal2007,paeckel2019}. In order to show the emergence of the second dark state $\ket{\rm D_e}$ in the effective Hamiltonian, we simulate the dynamics in Eq. \eqref{H_eff_evo} for an infinite system,  starting from the state $\ket{\rm U}$, until a fixed point is reached. In the regime associated with the exact dark state this dynamics always ends up in the state $\ket{\rm D}$. For sufficiently large values of $\Omega_2$, instead, the effective time-evolution converges towards the emergent dark state $\ket{\rm D_e}$, featuring a finite density of contagious sites $\varrho_\bullet$. Through a similar MPS algorithm, we have also studied the Lindblad dynamics \eqref{Lindblad} for an infinite system. For the model considered, this method  showed instabilities for large dimensions of the MPS. We have thus exploited a low-rank approximation of $\rho_t$, which may be regarded as an augmented mean-field description, and is expected to be valid for large enough $\Omega_2$. Our results show that the emergent dark state $\ket{\rm D_e}$ approached by the dynamics in Eq. \eqref{H_eff_evo}, is also the steady state of the average quantum dynamics [see Fig. \ref{fig:fig3}(c)].
\vspace{10pt}

\noindent {\bf Towards an experimental implementation.---} For the purpose of this work Hamiltonian \eqref{Hamiltonian} should be regarded as an idealized model. However, as we will briefly sketch in the following, one may indeed realize versions of it on current quantum simulator platforms based on Rydberg atoms \cite{bloch2012,labuhn2016,endres2016,bernien2017,keesling2019,browaeys2020,ebadi2020}. Here, the three states are represented by atomic Rydberg states which interact with a nearest-neighbor density-density interaction, parametrized by the matrix $V_{\alpha \beta}$ ($\alpha,\beta=\bullet,*,\circ$). The interaction Hamiltonian of a single atom with its left (${\rm L}$) and right $({\rm R})$ neighbor is then given by
\begin{equation*}
H_{\rm int}= \sum_{\alpha,\beta=\bullet,*,\circ}^3 V_{\alpha\beta}(n^\mathrm{(L)}_\beta+n^\mathrm{(R)}_\beta) n_\alpha.
\label{H_at}
\end{equation*}
The atoms are also driven by two lasers which couple the transitions $\ket{\bullet}\leftrightarrow\ket{*}$ (Rabi frequency $\Omega_1$) and $\ket{*}\leftrightarrow\ket{\circ}$ (Rabi frequency $\Omega_2$). By appropriately choosing the laser detunings and the coefficients of the interaction matrix $V_{\alpha \beta}$, and moving into a suitable interaction picture \cite{SM}, one obtains (in $1D$) the constrained Hamiltonian \cite{ates2007,sun2008,pohl2009,ates2012,jau2016,valado2016,marcuzzi2017,turner2018,lin2019,schecter2019,pancotti2020,sala2020} 
\begin{equation*}
\begin{split}
H_{\rm exp}&\approx \sum_k \Big[\Omega_{1}\left(1-n_{\circ}^{(k-1)}\right)\left(1-n_{\circ}^{(k+1)}\right) \lambda^{(k)}_1+\\
&+\Omega_{2}\left(n^{(k-1)}_{\bullet}+n_{\bullet}^{(k+1)}-2n^{(k-1)}_{\bullet}n_{\bullet}^{(k+1)}\right)\lambda^{(k)}_6\Big].
\end{split}
\end{equation*}
This is similar to the Hamiltonian \eqref{Hamiltonian}, with two differences: (i) a constraint on the transition $\ket{\bullet}\leftrightarrow\ket{*}$, which is not problematic, since it does not matter that the transition $\ket{\bullet}\leftrightarrow\ket{*}$ is switched off when the dark state $\ket{\rm D}$ is reached; (ii) a term proportional to $n_\bullet^{(k-1)}n_\bullet^{(k+1)}\lambda^{(k)}_6$. This term actually changes the qualitative behaviour, as we have tested with numerical simulations. Following Ref. \cite{wintermantel2020}, it can be eliminated by applying a further laser field with a detuning $2V_{\bullet \circ}$ and Rabi frequency $2\Omega_{2}$, on the $\ket{*}\leftrightarrow\ket{\circ}$ transition (see also discussion in Ref. \cite{nigmatullin2021}). This generates the counter term $2\Omega_{2}\sum_k n_\bullet^{(k-1)}n_\bullet^{(k+1)}\lambda^{(k)}_6$.
\vspace{10pt}

\noindent {\bf Discussion.---} We have investigated a novel type of nonequilibrium phase transition between two dark states, a trivial (classical) one and an emergent one. For the model considered, this transition appears to be present in all lattice dimensions. As we have shown, the detection of such an emergent state is possible by analyzing the spectral properties of the effective Hamiltonian. It is important to comment on why such a phenomenology cannot be observed in classical models. Here, a non-fluctuating state can only be a configuration state, as for example $\ket{\rm D}$, and thus can only be dark if it is an exact dark state for any system size. Furthermore, in classical settings, there is no effective coherent dynamics between jumps that could drive the system toward an emergent dark state. Finally, we note that, while dark states have already been investigated and explored in several quantum systems \cite{griessner2006,diehl2008,kraus2008,diehl2011}, the one we observe here is rather different in nature. Usually, dark states for quantum systems, even when showing entanglement and quantum correlations, are frustration-free dark states which are exact dark states for any system size. This means that their structure is such that they are eigenstates (with real eigenvalue) of each of the local terms in the effective Hamiltonian. In addition, they are annihilated by all jump operators. Instead, the dark state $\ket{\rm D_e}$ emerges as a collective property of the system, and is thus a real dark state only in the thermodynamic limit. This fact confers to it the robustness against perturbations which is associated with a genuine nonequilibrium phase. 
\vspace{10pt}

We  acknowledge  support  from the  “Wissenschaftler R\"uckkehrprogramm GSO/CZS” of the Carl-Zeiss-Stiftung and the German  Scholars Organization e.V., as well as through The Leverhulme Trust [Grant No.RPG-2018-181], and the Deutsche Forschungsgemeinschaft through SPP 1929 (GiRyd), Grant No. 428276754 and through Grant No. 435696605.

\bibliography{Reference}

\section*{Methods}
\subsection{Infinite-dimensional lattice}
In an infinite-dimensional lattice, each site has, as nearest neighbors, all the remaining ones forming the many-body system. As stated in the main text, this results in a constraint $\Pi_\bullet^k$ of the form given in Eq. \eqref{constrain-infinite}. Therefore, the dynamics of the average quantum state can be described through the quantum master equation \eqref{Lindblad}, with a Hamiltonian given by 
\begin{equation}
H\approx  \Omega_1 \sum_{k=1}^N\lambda_1^{(k)}+\frac{\Omega_2}{N}\sum_{k,h=1}^N n_\bullet^{(h)}\lambda_6^{(k)}\, .
\label{H_eff_infinite}
\end{equation}
The approximation is just due to the fact that we have added terms with $k=h$ in the double sum. In the limit $N\to\infty$, this does not constitute a problem, since these terms give rise to an intensive Hamiltonian contribution,  irrelevant for the dynamics of the system observables. 

In order to investigate the stationary behavior of the model, we focus on collective operators describing ``sample-average" properties. We consider
\begin{equation}
P_\alpha=\frac{1}{N}\sum_{k=1}^N n_\alpha^{(k)}\, ,\quad \mbox{with } \alpha=\bullet,*,\circ\, ,
\label{coll-op1}
\end{equation}
representing the density of sites in the different states, as well as 
\begin{equation}
G_\alpha=\frac{1}{N}\sum_{k=1}^N \lambda_\alpha^{(k)}\, ,\quad \mbox{ for } \alpha=1,2,4,5,6,7\, .
\label{coll-op2}
\end{equation}
We stress here that we are using the Gell-Mann matrices $\lambda_\alpha$ for off-diagonal single-site observables, while we use the projectors $n_\alpha$ for the diagonal ones. 

\subsubsection{Average Lindblad Dynamics}
The collective operators of Eqs. \eqref{coll-op1}-\eqref{coll-op2} behave, when considering the state $\ket{\rm U}$ and the large $N$ limit, as ``classical" scalar quantities equal to their expectation value on the state \cite{lanford1969}. This means that we have $G_\alpha\to \Lambda_\alpha=\braket{\lambda_\alpha}$ as well as $P_\alpha\to \varrho_\alpha=\braket{n_\alpha}$, where we have also exploited the translation invariance of the state. For the infinite-dimensional lattice, the dynamics of these operators is captured by the following non-linear differential equations \cite{benatti2018} ($\partial_t:=d/dt$)
\begin{eqnarray}
\partial_t \Lambda_1 &=&\Omega_{2}\varrho_\bullet \Lambda_5-\Omega_{2}\Lambda_2\Lambda_6-\frac{\gamma}{2}\Lambda_1\, ,
\nonumber\\
\partial_t \Lambda_2 &=&-2\Omega_{1}(\varrho_\bullet-\varrho_*)-\Omega_{2}\varrho_\bullet \Lambda_4+\Omega_{2}\Lambda_1\Lambda_6-\frac{\gamma}{2}\Lambda_2 \, ,
\nonumber\\
\partial_t \Lambda_4 &=& -\Omega_{1}\Lambda_7+\Omega_{2}\varrho_\bullet \Lambda_2-\Omega_{2}\Lambda_5\Lambda_6\, ,
\nonumber\\
\partial_t \Lambda_5 &=& \Omega_{1}\Lambda_6-\Omega_{2}\varrho_\bullet \Lambda_1+\Omega_{2}\Lambda_4\Lambda_6\, ,
\nonumber\\
\partial_t \Lambda_6 &=& -\Omega_{1}\Lambda_5-\frac{\gamma}{2}\Lambda_6\, ,\label{mf-eqs}\\
\partial_t \Lambda_7 &=& \Omega_{1}\Lambda_4-2\Omega_{2}\varrho_\bullet(\varrho_*-\varrho_\circ)-\frac{\gamma}{2}\Lambda_7\, ,
\nonumber\\
\partial_t \varrho_\bullet&=&\Omega_{1}\Lambda_2\, ,
\nonumber\\
\partial_t \varrho_*&=&-\Omega_{1}\Lambda_2+\Omega_{2}\varrho_\bullet \Lambda_7-\gamma\varrho_*\, ,
\nonumber\\
\partial_t \varrho_\circ&=&-\Omega_{2}\varrho_\bullet \Lambda_7+\gamma\varrho_*\, .
\nonumber
\end{eqnarray}
To find the possible stationary states, we set to zero all the time-derivatives above. In this way, we find the following three solutions: 
\begin{center}
\begin{tabular}{cccc}
$\varrho_\bullet=0$,& $\varrho_\circ=1$,& $\Lambda_4= 0$;\\
$\varrho_\bullet=\frac{1}{2}\left[1-\sqrt{1-x^2}\right]$,& $\varrho_\circ=\frac{1}{2}\left[1+\sqrt{1-x^2}\right]$,& $\Lambda_4=-x$;\\
$\varrho_\bullet=\frac{1}{2}\left[1+\sqrt{1-x^2}\right]$,& $\varrho_\circ=\frac{1}{2}\left[1-\sqrt{1-x^2}\right]$,& $\Lambda_4=-x$;
\end{tabular}
\end{center}
with $x=2\Omega_1/\Omega_2$ and all other collective quantities being zero. The last two solutions are physically meaningful only for $|x|\le1$. This shows the existence of a critical rate $\Omega_2^{\rm c}=2\Omega_1$ below which the unique admissible solution is the first one, which is the exact dark state $\ket{\rm D}$. When $\Omega_2\ge\Omega_2^{\rm c}$ ($|x|\le1$), two further solutions become possible. The second one turns out to be unstable so that it is never dynamically approached. The third, instead, corresponds to a (stable) emergent stationary state with a finite density of sites in $\ket{\bullet}$. This state is pure and --- concerning the computation of local observables or of collective operators such as those introduced before --- can be represented as the state 
$$
\ket{\rm D_e}=\bigotimes_{k=1}^N \left(\alpha_+\ket{\bullet}-\alpha_-\ket{\circ}\right)^{(k)}\, ,
$$
with $\alpha_\pm=\sqrt{2}^{-1}\sqrt{1\pm\sqrt{1-x^2}}$. However, due to the all-to-all coupling in the Hamiltonian of Eq. \eqref{H_eff_infinite}, the state $\ket{\rm D_e}$ can develop weak long-range correlations \cite{benatti2018}. In order to verify which stationary state is dynamically approached when the system is initially in $\ket{\rm U}$, we have performed a numerical integration of Eqs. \eqref{mf-eqs}, up to large times for which the state is  stationary. These results are presented in Fig. \ref{fig:fig2}(a-b) in the main text. 

\subsubsection{Exact diagonalization of the effective Hamiltonian}
To show that the emergent dark state is approached through the dynamics of Eq. \eqref{H_eff_evo}, we have performed exact diagonalization of the non-Hermitean Hamiltonian $H_{\rm eff}$. The idea is that since the emergent stationary state $\ket{\rm D_e}$ is pure, it must be, in the $N\to\infty$ limit, an eigenvector of $H_{\rm eff}$ associated with a second eigenvalue which converges to zero in imaginary part. In this way, the rate of escaping from $\ket{\rm D_e}$ would be zero and the dynamics in Eq. \eqref{H_eff_evo} would have this state as a possible fixed point. Since we start from the fully symmetric state $\ket{\rm U}$, and since the average Lindblad dynamics is fully symmetric (invariant under any permutation of the three-level units), in order to detect the emergence of the second dark state $\ket{\rm D_e}$, it is sufficient to diagonalize the effective Hamiltonian in its fully symmetric subspace.  To do so, we have developed a representation of such an operator in the space of vectors which are invariant under any permutation of two subsystems. Such a subspace can be constructed, for instance, by acting on the fully symmetric vector $\ket{\rm U}$ with permutation invariant operators. The procedure to obtain such a representation of the effective Hamiltonian is presented in Ref. \cite{SM}.

\subsection*{One-dimensional lattice: simulations with matrix product states}
\subsubsection*{Effective Hamiltonian dynamics}
For the one-dimensional lattice, to investigate the transition from the dark state $\ket{\rm D}$ to the dark state $\ket{\rm D_e}$, we use a time-evolving-block-decimation algorithm (TEBD) which exploits the translation invariance of the model to address the system in the infinite one-dimensional lattice limit (iTEBD), see Refs. \cite{vidal2003,Vidal2004,vidal2007,paeckel2019} for details. Since we cannot perform exact diagonalization of $H_{\rm eff }$, we analyze the fixed point obtained by running the dynamics in Eq. \eqref{H_eff_evo}, starting from state $\ket{\rm U}$, for sufficiently large times. 

The formal integration of Eq. \eqref{H_eff_evo} gives the solution 
$$
\ket{\psi_t}=\frac{W_t\ket{\psi}}{\|W_t\ket{\psi}\|}\, ,
$$ 
where the time-propagator is given by $W_{t}=e^{-{\rm i}{t} H_{\rm eff}}$. In order to simulate such dynamics by means of matrix product state (MPS) methods, we first discretize time. We take a sufficiently small time step $dt$ and express the infinitesimal propagator $W_{dt}$ through the second-order Trotter decomposition
$$
W_{dt}\approx e^{-{\rm i}{dt}/2 H_{\rm ss}}e^{-{\rm i}{dt}/2 H_{\rm o}}e^{-{\rm i}{dt} H_{\rm e}}e^{-{\rm i}{dt}/2 H_{\rm o}}e^{-{\rm i}{dt}/2 H_{\rm ss}}\, ,
$$
where $H_{\rm ss}=\sum_k (\Omega_1\lambda_1^{(k)}-{\rm i\gamma}/2 n_{*}^{(k)})$ is the single-site component of the effective Hamiltonian, while $H_{\rm o/e}$ contain the nearest-neighbor interaction terms for odd and even bonds, respectively. 

We perform time updates for the MPS representation of the state $\ket{\psi_t}$, using a standard iTEBD method \cite{vidal2007}. After the application of the single-site gates in $e^{-{\rm i}{dt}/2 H_{\rm ss}}$, which are not unitary operators, we perform $N_{\rm corr}$ repetitions of a trivial time-evolution on both even and on odd bonds. This is done by applying two-body gates equal to the identity operator at the aim of keeping the MPS in canonical form. After each application of a two-body gate, we perform a singular value decomposition and a truncation step. We take only the $\chi_{\rm max}$ largest singular values, as long as these are larger than a truncation error $\varepsilon_{\rm trunc}$. We have observed that it is beneficial to have at least $N_{\rm corr}=1$. This also relates to the fact that, after the application of the non-unitary gates, the entanglement in the MPS can decrease and the trivial evolution step can reduce the bond dimension while preparing the MPS for the two-body evolution. However, we also noticed that having $N_{\rm corr}>1$ does not lead to further improvements. We mainly considered $N_{\rm corr}=2$.

Concerning the truncation error $\varepsilon_{\rm trunc}$, this does not seem to be relevant in the phase where the dark state $\ket{\rm D_e}$ appears. Its precise value affects instead the transient dynamics in the phase related to state $\ket{\rm D}$. However, this is not expected to have impact on its fixed point. We explored values for the truncation error ranging from $\varepsilon_{\rm trunc}=10^{-10}$ to $\varepsilon_{\rm trunc}=10^{-14}$. 

The results obtained with this iTEBD method are reported in Fig. \ref{fig:fig3}(c). We furthermore mention here that the representative trajectories appearing in Fig. \ref{fig:fig1}(c) have been obtained by simulating the stochastic quantum dynamics through MPSs and a TEBD method \cite{Vidal2004}, for a finite chain with open boundary conditions.

\subsubsection{Average Lindblad dynamics}
We have also simulated the open quantum dynamics of the average state $\rho_t$ --- which obeys the quantum master equation \eqref{Lindblad} --- through MPS methods (see e.g. also Ref. \cite{carollo2019} or Ref. \cite{kshetrimayum2017} for an application to $2D$ dissipative systems). The idea is to represent the density matrix $\rho_t$ as a vector $\ket{\rho_t}$ in an enlarged Hilbert space. This is achieved through the mapping 
$$
\rho=\sum_{\vec{\alpha},\vec{\beta}}r_{\vec{\alpha}\vec{\beta}}\ket{\vec{\alpha}}\!\bra{\vec{\beta}}\to \ket{\rho}=\sum_{\vec{\alpha},\vec{\beta}}r_{\vec{\alpha}\vec{\beta}}\bigotimes_{k=1}^N \left(\ket{\alpha_k}\otimes \ket{\beta_k}\right)^{(k)}\, ,
$$
where $\vec{\alpha}=(\alpha_1,\alpha_2,\dots \alpha_N)$ is a vector specifying the single-particle state of the different sites for the many-body state $\ket{\vec{\alpha}}$ and  $r_{\vec{\alpha}\vec{\beta}}=\bra{\vec{\alpha}}\rho\ket{\vec{\beta}}$. 

In this representation, our Lindblad generator $\mathcal{L}$ becomes the following matrix 
\begin{equation}
\begin{split}
L&=\gamma \sum_{k=1}^N \left(J^{(k)}-\frac{1}{2} \hat{n}_{*,{\rm I}}^{(k)}-\frac{1}{2}\hat{n}_{*,{\rm II}}^{(k)}\right)\\
&+\sum_{k=1}^{N}\Omega_1\left(-{\rm i} \lambda_{1,{\rm I}}^{(k)} +{\rm i}\lambda_{1,{\rm II}}^{(k)}\right)+\, \\
&-{\rm i}\sum_{k=1}^{N-1}\frac{\Omega_2}{2}\left(n_{\bullet,{\rm I}}^{(k-1)}+n_{\bullet,{\rm I}}^{(k+1)}\right)\lambda_{6,{\rm I}}^{(k)}+\\
&+{\rm i}\sum_{k=1}^{N-1}\frac{\Omega_2}{2}\left(n_{\bullet,{\rm II}}^{(k-1)}+n_{\bullet,{\rm II}}^{(k+1)}\right)\lambda_{6,{\rm II}}^{(k)}\, ,
\end{split}
\end{equation}
where $J=J_-\otimes J_-$, $\hat{n}_{\alpha,{\rm I}}=n_\alpha\otimes {\bf 1}_3$, $\hat{n}_{\alpha,{\rm II}}= {\bf 1}_3\otimes n_\alpha$, and ${\bf 1}_3$ is the three-dimensional identity matrix. In addition we have $\lambda_{\alpha,{\rm I}}=\lambda_\alpha \otimes {\bf 1}_3$ and $\lambda_{\alpha,{\rm II}}={\bf 1}_3\otimes \lambda_\alpha^{T} $. 

The dynamics is implemented through the equation
$$
\frac{d}{dt} \ket{\rho_t}=L\ket{\rho_t}\, . 
$$
This can be formally integrated to obtain $\ket{\rho_t}=e^{t L}\ket{\rho}$, and, since the matrix $L$ contains at most two-site interactions, this evolution can be approximated with the same iTEBD method discussed in the previous section. As initial state, we consider the density matrix $\rho=\ket{\rm U}\!\bra{\rm U}$, represented through the vector 
$$
\ket{\rho}=\bigotimes_{k=1}^N \left(\ket{\bullet}\!\ket{\bullet}\right)^{(k)}\, .
$$

The main difference with the algorithm used to simulate the time-evolution under $H_{\rm eff}$, lies in the way expectation values are computed. In this case, to compute the expectation value of a quantum observable, we first need to define a vector representation of the identity operator in the enlarged Hilbert space. This is given by 
$$
\ket{\rm id}=\bigotimes_{k=1}^N\ket{{\rm id}_1}^{(k)}\, ,
$$
where $\ket{{\rm id}_1}=\ket{\bullet}\!\ket{\bullet}+\ket{*}\!\ket{*}+\ket{\circ}\!\ket{\circ}$ is the vector representation of the single-site identity operator. Expectation values are then computed as
$$
\langle O^{(k)}\rangle_t=\bra{\rm id}O^{(k)}_{\rm L}\ket{\rho_t}\, ,
$$
with $O_{\rm L}=O\otimes {\bf 1}_3$. 

In the subcritical region --- the one associated with the exact dark state $\ket{\rm D}$ ---  we have observed a good convergence of iTEBD results when increasing the bond dimension $\chi_{\rm max}$ of the MPS representation of $\ket{\rho_t}$. On the other hand, for large values of $\Omega_2/\Omega_1$, for which the system is expected to belong to the emergent dark state phase, it was not possible to obtain convergence of the iTEBD results, as the algorithm showed instabilities. 

For $\Omega_2\gg \Omega_1$, where the system features values of $\varrho_\bullet$ close to one, the state of the system is close to the product state with all sites in $\ket{\bullet}$. In this regime, far from the critical point, augmented product-ansatz descriptions --- also known as augmented mean-field description --- able to account for the weak short-ranged correlations in the state are expected to capture the relevant properties of the state (see e.g. Ref. \cite{weimer2015} for an application to open quantum systems). In our numerical simulations, we have noted that the curves obtained with small bond dimensions ($\chi_{\rm max}=4,8$) are stable and provide a similar prediction for the steady-state value of $\varrho_\bullet$. Since even with a very small bond dimension, the MPS ansatz can account for short-ranged correlations beyond mean-field theories, we can regard such results as obtained from an augmented product-state description. The data obtained for these small bond dimensions are plotted in Fig. \ref{fig:fig3}(c). These are in agreement with the results obtained for the effective Hamiltonian. This suggests that the steady state of the Lindblad dynamics is indeed the emergent dark state of $H_{\rm eff}$ in the supercritical region --- the one associated with the emergent dark state $\ket{\rm D_e}$.

\setcounter{equation}{0}
\setcounter{figure}{0}
\setcounter{table}{0}
\makeatletter
\renewcommand{\theequation}{S\arabic{equation}}
\renewcommand{\thefigure}{S\arabic{figure}}
\makeatletter
\renewcommand{\theequation}{S\arabic{equation}}
\renewcommand{\thefigure}{S\arabic{figure}}

\onecolumngrid
\newpage

\begin{center}
{\Large SUPPLEMENTARY INFORMATION}
\end{center}
\begin{center}
\vspace{0.8cm}
{\Large Nonequilibrium dark space phase transition}
\end{center}
\begin{center}
Federico Carollo$^{1}$ and Igor Lesanovsky$^{1,2}$
\end{center}
\begin{center}
$^1${\em Institut f\"ur Theoretische Physik, Universit\"at T\"ubingen,}\\
{\em Auf der Morgenstelle 14, 72076 T\"ubingen, Germany}\\
$^2$ {\em School of Physics and Astronomy and Centre for the Mathematics}\\
{\em and Theoretical Physics of Quantum Non-Equilibrium Systems,}\\
{\em The University of Nottingham, Nottingham, NG7 2RD, United Kingdom}
\end{center}

\subsection*{Representation of the effective Hamiltonian in the fully symmetric subspace}
To develop a representation of the Hamiltonian $H_{\rm eff}$ for the infinite-dimensional lattice in the fully symmetric subspace, it is convenient to use all the relevant Gell-Mann matrices $\lambda_i$, for $i=1,2,\dots 8$. Their algebraic structure is encoded in the commutation relations
\begin{equation}
[\lambda_i,\lambda_j]=2i\sum_{\ell=1}^8\epsilon_{ij\ell} \lambda_\ell\, ;
\label{gell-comm}
\end{equation}
importantly, $\epsilon_{ij\ell}$ is an anti-symmetric tensor. We further define the matrices $f_i=\frac{1}{2}\lambda_i$. 

In order to efficiently represent the Hamiltonian, we need to derive a representation for collective operators constructed as 
\begin{equation}
F_i=\sum_{k=1}^Nf_i^{(k)}\, ,
\label{F-coll-op}
\end{equation}
as acting on fully symmetric states. For these operators, the commutation relations are inherited from Eq.~\eqref{gell-comm}
\begin{equation}
[F_i,F_j]=\sum_{k,h=1}^N\frac{1}{4}\left[\lambda_i^{(k)},\lambda_{j}^{(h)}\right]=\sum_{k=1}^N\frac{1}{4}\left[\lambda_i^{(k)},\lambda_{j}^{(k)}\right]=i\sum_{\ell=1}^8\epsilon_{ij\ell}F_\ell\, .
\label{coll-comm}
\end{equation}
Exploiting the anti-symmetric property of the tensor $\epsilon_{ijk}$, it is possible to show that the operator
$$
C=\sum_{i=1}^8F_i^2\, ,
$$
is the (quadratic) Casimir operator for the algebra formed by the operators $F_i$, as it commutes with each of them. Indeed, one has that 
$$
[C,F_j]=\sum_{i=1}^8\left(F_i[F_i,F_j]+[F_i,F_j]F_i\right)=i\sum_{i,\ell=1}^8\epsilon_{ij\ell}\left(F_iF_\ell +F_\ell F_i\right)\, ,
$$ 
and looking at the last relation in the above equation, one sees that the term in the round brackets is symmetric with respect to exchange of $i\leftrightarrow\ell$, while the term $\epsilon_{ij\ell}$ is anti-symmetric. As such the total double sum is zero. 

Because of this, the value of the Casimir operator in the fully symmetry sector could be computed by considering any state in the subspace, for instance also the state 
$$
\ket{\rm U}=\bigotimes_{k=1}^N\ket{\bullet}^{(k)}\, ,\qquad \mbox{ where }\qquad \ket{\bullet}=\begin{pmatrix}
1\\0\\0
\end{pmatrix}\, .
$$
\vspace{10pt}

To make progress in finding a representation for the operators in Eqs. \eqref{F-coll-op}, we now introduce a new set of operators, forming the so-called Cartan-Weyl basis, which are constructed from the $F_i$ as follows:
\begin{equation}
I_\pm=F_1\pm i F_2\, , \quad  I_3=F_3\, , \quad  V_{\pm} =F_4\pm i F_5\, ,\quad U_\pm =F_6\pm i F_7\, ,\quad Y=\frac{2}{\sqrt{3}}F_8\, .
\label{C-W-operators}
\end{equation}
The commutation relations between these operators can be computed from the fundamental ones in Eq. \eqref{coll-comm}. We notice that $[I_3,Y]=0$, so that these two operators can be simultaneously diagonalized. We define the normalized eigenstates of both $I_3$ and $Y$ to be $\ket{t,y}$. In particular, we have
$$
I_3\ket{t,y}=t\ket{t,y}\, ,\qquad Y\ket{t,y}=y\ket{t,y}\, .
$$
Basically, we have that $t=(N_{\bullet}-N_{*})/2$, where $N_\alpha$ denotes the total number of particles in state $\ket{\alpha}$, with $\alpha=\bullet,*,\circ$, and that $y=N_\bullet/3+N_{*}/3-2N_{\circ}/3$. 
As such,
$-N/2\le t\le N/2$ and $-2N/3\le y\le N/3$. This set of states forms a complete orthonormal basis for the subspace of interest. 

For the vectors $\ket{t,y}$, the operators $I_\pm,V_\pm,U_\pm$ act as ladder operators. As a simple example of this fact, lets consider the state $\ket{\rm U}$ which can be written as $\ket{\rm U}=\ket{N/2,N/3}$. This is annihilated by the action of $I_+$, which wants to bring a particle from $\ket{*}$  to $\ket{\bullet}$. On the other hand, we have 
$$
I_-\ket{N/2,N/3}=\alpha  \ket{N/2-1,N/3}\, ,
$$
where $\alpha$ has to be determined through the norm of $I_-\ket{N/2,N/3}$. Using that $I_+$ annihilates $\ket{\rm U}$ we have that 
\begin{equation}
\begin{split}
\bra{N/2,N/3}I_+I_-\ket{N/2,N/3}&=\bra{N/2,N/3}[I_+,I_-]\ket{N/2,N/3}=N\, .
\end{split}
\label{example}
\end{equation}
So in this case, $\alpha=\sqrt{N}$. 
\vspace{10pt}

In order to find the appropriate values of $\alpha$ for any state onto which $I_-$ is acting, i.e. the value of $\alpha$ for generic 
$$
I_-\ket{t,y}=\alpha \ket{t-1,y}\, ,
$$
we need to develop a systematic approach.  Lets focus for the moment on the subalgebra formed by $I_\pm$ and $I_3$. These operators are made of the sum of $3\times3$ matrices which basically represent the spin-$1/2$ algebra (the algebra of Pauli matrices) embedded into a largest ($3\times3$) space. This can also be seen by noticing that $I_\pm$ and $I_3$ have among them the same commutation relations of spin-$1/2$ operators. We then compute  
$$
I_+I_-=F_1^2+F_2^2+i[F_2,F_1]=\left[F_1^2+F_2^2+F_3^2\right]-F_3(F_3-1)\, .
$$
The operator in the square brackets is a ``partial" Casimir operator, which we call $C_{12}=\sum_{i=1}^3 F_i^2$, since it commutes with the subalgebra formed  by the $F_i$, with $i=1,2,3$. As such, $C_{12}$ also commutes with the $I_\pm$. As apparent from Eq. \eqref{example}, it is important to understand the value of this operator on different states in order to be able to compute the norm of vectors such as $I_-\ket{t,y}$.

Since $C_{12}$ commutes with $I_{\pm}$ and that $I_\pm$ does not modify $y$, to find the value of this partial Casimir operator $C_{12}$ in the different subspaces (labelled by the value of $y$), it is convenient to find a simple state to compute its expectation value.  In particular, we look for the one which has the maximum value of $t$ allowed, for a fixed value of $y$ (and given that we have only $N$ particles). We call $N_{\bullet *}$ the number of particles which are in state $\ket{\bullet}$ or $\ket{*}$. We thus have $N_{\bullet *}+N_\circ=N$ and also that, given a value of $\bar{y}$ we have 
$$
\bar{y}=\frac{N_{\bullet *}-2N_\circ}{3}\, , \longrightarrow  N_{\bullet*}=\bar{y}+\frac{2N}{3}\, .
$$
Then, if all the particles in the subspace formed by the states $\ket{\bullet},\ket{*}$ are found in the state $\ket{\bullet}$ we have the maximum 
$$
t_{{\rm max}|\bar{y}}=\frac{N_{\bullet *}}{2}=\frac{\bar{y}+\frac{2N}{3}}{2}\, .
$$
We can thus find the value of the partial Casimir operator $C_{12}$, at fixed $\bar{y}$, which is given by 
$$
\bra{t_{{\rm max}|\bar{y}},\bar{y}}C_{12}\ket{t_{{\rm max}|\bar{y}},\bar{y}}=\frac{N_{\bullet *}}{2}\left(\frac{N_{\bullet*}}{2}+1\right)=\left(\frac{\bar{y}+\frac{2N}{3}}{2}\right)\left(\frac{\bar{y}+\frac{2N}{3}}{2}+1\right)\, .
$$
With this result, we can now find all matrix elements for the operator $I_-$, in the fully symmetric subspace. Indeed, we use that 
\begin{equation}
I_-\ket{t,y}=\alpha_{t,y} \ket{t-1,y}\, ,
\label{ladder1}
\end{equation}
where $\alpha_{t,y}^2$ is given by 
$$
\alpha_{t,y}^2=\bra{t,y}I_+I_-\ket{t,y}=\bra{t,y}C_{12}-I_3(I_3-1)\ket{t,y}=t_{{\rm max}|y}\left(t_{{\rm max}|y}+1\right)-t\left(t-1\right)\, .
$$

Now we use an analogous procedure to represent the ladder operator $U_-$.  The action of $U_-$ on a state consists in taking one particle from state $\ket{*}$ and bringing it to state $\ket{\circ}$. This means that the action of $U_-$ reduces $y$ by $1$ but also increments $t$ by $1/2$. We thus have the relation 
\begin{equation}
U_-\ket{t,y}=\beta_{t,y} \ket{t+\frac{1}{2},y-1}\, ;
\label{ladder2}
\end{equation}
Similarly to what done before, the factor $\beta^2_{t,y}$ can be written as  
$$
\beta_{t,y}^2=\bra{t,y}U_+ U_-\ket{t,y}=\bra{t,y}\left[C_{23}-\tilde{F}_8\left(\tilde{F}_8-1\right)\right]\ket{t,y}\, ,
$$
where 
$$
\tilde{F}_8=\sum_{k=1}^N \frac{1}{4}\left(\sqrt{3}\lambda_8-\lambda_3\right)^{(k)}\, , 
$$
while $C_{23}=F_6^2+F^2_7+\tilde{F}_8^2$ is the partial Casimir operator, now for the subspace generated by $\ket{*},\ket{\circ}$. We note that $\tilde{F}_8=\frac{3}{4}Y-\frac{1}{2}I_3$. The task is to find a way to compute the value of this partial Casimir operator on a simple reference state. We proceed as follows. Lets consider a generic state $\ket{t,y}$, and remember that we have $N$ particles. We thus can find
\begin{equation}
\begin{split}
2t=N_{\bullet}-N_{*}\, \quad y=N_\bullet+N_{*}-\frac{2N}{3}\, .
\end{split}
\end{equation}
Inverting the above relations, we can find an expression for $N_\bullet,N_{*}$ (and thus also for $N_\circ$ if necessary) as a function of $t,y$. We have 
$$
N_\bullet=t+\frac{y}{2}+\frac{N}{3}\, , \qquad N_{*}=\frac{y}{2}-t+\frac{N}{3}\, .
$$
As already said, the action of the operator $U_-$ is that of taking one particle from $\ket{*}$ and bringing it to $\ket{\circ}$. As such, without changing the value of the partial Casimir operator (since $[U_-,C_{23}]=0$), this state is connected by the repeated action of $U_-$ to the state with $N_\bullet=t+\frac{y}{2}+\frac{N}{3}$ as before, $N_{*}=0$ and $N_\circ=N-N_\bullet$. Such a state is given by 
$$
\ket{\varphi}=\ket{\frac{t}{2}+\frac{y}{4}+\frac{N}{6},t+\frac{y}{2}-\frac{N}{3}}\, ,
$$
and, from this, the partial Casimir operator $C_{23}$ can be computed solely as a function of $t,y$ of the original state. 
In particular, we have 
$$
\bra{\varphi}C_{23}\ket{\varphi}=\frac{N_\circ}{2}\left(\frac{N_\circ}{2}+1\right)=\frac{1}{2}\left(\frac{2N}{3}-t-\frac{y}{2}\right)\left[\frac{1}{2}\left(\frac{2N}{3}-t-\frac{y}{2}\right)+1\right]\, ,
$$
and, therefore, 
$$
\beta_{t,y}^2=C_{23}-\left[\frac{3}{4}y-\frac{t}{2}\right]\left[\frac{3}{4}y-\frac{t}{2}-1\right]\, . $$

\vspace{10pt}
This procedure allows us to represent the effective Hamiltonian in Eq. \eqref{H_eff_infinite}, in the fully symmetric subspace spanned by all vectors $\ket{t,y}$. Using relations \eqref{ladder1} and \eqref{ladder2}, we indeed have  
$$
I_-=\sum_{t,y}\alpha_{t,y}\ket{t-1,y}\bra{t,y}\, ,
$$
and  
$$
U_-=\sum_{t,y}\beta_{t,y}\ket{t+\frac{1}{2},y-1}\bra{t,y}\, .
$$
We can use this to represent the effective Hamiltonian noticing that 
$$
\sum_{k=1}^N \lambda_1^{(k)}=I_-+I_+\, ,
$$
and also that 
$$
\sum_{k=1}^N \lambda_6^{(k)}=U_-+U_+\, .
$$
Furthermore, we also have 
$$
\sum_{k=1}^Nn_\bullet^{(k)}=I_3+\frac{Y}{2}+\frac{N}{3}\, ,
$$
and
$$
\sum_{k=1}^Nn_{*}^{(k)}=-I_3+\frac{Y}{2}+\frac{N}{3}\, .
$$
We have used this representation of the Hamiltonian to produce numerical data for the plots presented in Fig. \ref{fig:fig2}(c-d) as well as those shown in Fig. \ref{fig:fig_gap_SM}.

\newpage

\subsection*{Additional numerical results for the infinite-dimensional lattice}
\begin{figure*}[h]
    \centering
    \includegraphics[width=\linewidth]{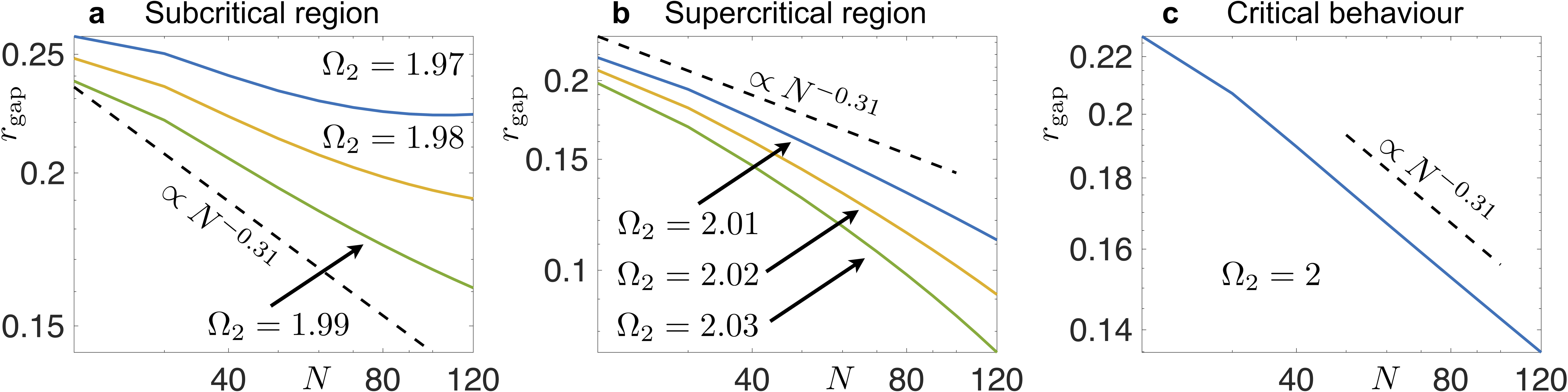}
    \caption{\textbf{Gap of the effective Hamiltonian for the infinite-dimensional lattice}. Log-log plots of the gap $r_{\rm gap}$ of $H_{\rm eff}$ as a function of $N$ in different parameter regimes for $\Omega_1=1$ (in units of $\gamma$). (a) In the subcritical region, identified by values of $\Omega_2<2\Omega_1$, the gap shows a tendency to saturate to a finite value. This is manifesting in the log-log plot via curves which are concave up. (b) In the supercritical region, instead we have that curves are concave down which indicates an exponential decay of the gap as a function of $N$. (c) At criticality, $\Omega_2=2\Omega_1$, the gap tends to vanish with a power-law behaviour, with an exponent approximately equal to $-0.31$. In all panels, the dashed line, proportional to $N^{-0.31}$, is shown for comparison.  }
    \label{fig:fig_gap_SM}
\end{figure*}

In this section, we present data showing that the gap of the $H_{\rm eff}$ Hamiltonian remains finite in the region $\Omega_2<2\Omega_1$ while it tends to zero for values of $\Omega_2>2\Omega_1$. At the critical point $\Omega_2=2\Omega_1$, we have instead a power-law decay of the gap with an exponent close to $0.31$. All of this is shown in Fig. \ref{fig:fig_gap_SM}.

\vspace{20pt}

\newpage

\subsection*{Practical implementation of the constraint}
In this section we provide details on the discussion about the possibility of implementing the constrained system Hamiltonian in experiments reported in the main text.

We consider an experimental setting involving several three-level Rydberg atoms, arranged in a $1D$ array. We take the following Hamiltonian
\begin{equation}
H_{\rm exp}=\Omega_1 \sum_{k} \lambda_1^{(k)}+\Omega_2 \sum_{k} \lambda_6^{(k)}+H_{\rm diag}\, ,
\label{H_exp}
\end{equation}
whose first two terms represent two laser drivings, while $H_{\rm diag}=\sum_k H_{\rm at}^{(k)}$, where $H_{\rm at}^{(k)}$ contains the interaction energy of a single atom associated with its configuration and that of its neighbors (the left one, ${\rm L}$, and the right one, ${\rm R}$) as well as laser detuning terms. For a reference atom, we have 
\begin{eqnarray}
H_{\rm at}=H_{\rm int }+H_{\rm det }\, , \qquad  \mbox{ with } \quad H_{\rm int }=\sum_{\alpha,\beta=\bullet,*,\circ}^3 \!\!V_{\alpha\beta}(n^\mathrm{(L)}_\beta+n^\mathrm{(R)}_\beta) n_\alpha\, \, ,\quad H_{\rm det }=\sum_{\alpha=\bullet,*,\circ }^3 \!\!h_\alpha n_\alpha \, .
\end{eqnarray}
here, $h_\alpha$ are the detunings while the symmetric matrix $V_{\alpha \beta}$ encodes the state-dependent atomic interactions. 

The main idea to obtain the desired constraint consists in ``rotating" the system Hamiltonian $H_{\rm exp}$ into an interaction picture obtained by subtracting to the time-evolution, the unitary operator 
$$
U_t=\prod_k e^{-{\rm i}t H_{\rm at}^{(k)}}
$$
In this frame, the Hamiltonian $H_{\rm exp}$ transforms into
\begin{equation}
H_{\rm exp}'=U_t^\dagger \left(\sum_k \Omega_1\lambda_1^{(k)}+\Omega_2\lambda_6^{(k)}\right)U_t\, .
\label{SM:H_exp}
\end{equation}
To explicitly write down such an operator,  we need to evaluate the action of $e^{{\rm i}t H_{\rm at}}$ on the different states $\ket{\circ},\ket{*},\ket{\bullet}$. This is indeed enough to understand how the off-diagonal matrices $\lambda_1$ and $\lambda_6$ transform in the rotating frame. We will also apply a rotating wave approximation. Different choices of $h_\alpha$ and $V_{\alpha\beta}$ can thus give rise to different constraints. The first step is thus to compute the following expression ($\alpha=\bullet,*,\circ$):
\begin{eqnarray}
\exp(iH_\mathrm{at} t) |\alpha\rangle&=&\exp\left(it \left[h_\alpha + \sum_{\beta=\bullet,*,\circ}^3 V_{\alpha\beta}(n^{(\mathrm{L})}_\beta+n^{(\mathrm{R})}_\beta)\right] \right) |\alpha\rangle=\nonumber\\
&=&\exp\left(it h_\alpha\right)  \left[1-\sum_\beta n^{(\mathrm{L})}_\beta\left[1-e^{it V_{\alpha\beta}}\right] \right] \left[1-\sum_{\beta'} n^{(\mathrm{R})}_{\beta'}\left[1-e^{it V_{\alpha\beta'}}\right] \right]  |\alpha\rangle\, .
\end{eqnarray}
The aim is to constrain the transition between states $\ket{\circ}$ and $\ket{*}$. To this end, we can choose $h_{\bullet}=h_{*}=0$, as well as $V_{\alpha\beta}$ with non-zero components only given by $V_{\bullet\circ}=V_{\circ\bullet}$. In this way, by also fixing $h_\circ=-V_{\circ\bullet}$ and neglecting oscillating terms, we find 
\begin{equation*}
    e^{iH_{\rm at }t }\ket{*}=\ket{*}\, \qquad e^{iH_{\rm at }t }\ket{\circ}\approx \ket{\circ}\left[n_\bullet^{({\rm L})}+n_\bullet^{({\rm R})}-2n_\bullet^{({\rm L})}n_\bullet^{({\rm R})}\right]\, ,
\end{equation*}
and
$$
e^{iH_{\rm at }t }\ket{\bullet}\approx \ket{\bullet}\left[\left(1-n_\circ^{({\rm L})}\right)+\left(1-n_\circ^{({\rm R})}\right)\right]\, ,
$$

Using these results for the Hamiltonian in Eq. \eqref{SM:H_exp} leads to the many-body operator
\begin{equation}
H_{\rm exp}'\approx \sum_k \Big[\Omega_{1}\left(1-n_{\circ}^{(k-1)}\right)\left(1-n_{\circ}^{(k+1)}\right) \lambda^{(k)}_1
+\Omega_{2}\left(n^{(k-1)}_{\bullet}+n_{\bullet}^{(k+1)}-2n^{(k-1)}_{\bullet}n_{\bullet}^{(k+1)}\right)\lambda^{(k)}_6\Big]\, ,
\end{equation}
which is the one reported in the main text. 
\vspace{10pt}

\end{document}